\documentclass[10pt]{amsart}

\usepackage[a4paper,headheight=0.3cm]{geometry}
\usepackage{fancyhdr}
\usepackage{url}
\usepackage{enumerate}
\usepackage{amsfonts}
\usepackage{amsmath}
\usepackage{tikz}
\usepackage{comment}
\usepackage{bm}
\usepackage[pdftitle={A Robust Bayesian Approach to Modelling Epistemic Uncertainty in Common-Cause Failure Models},%
pdfauthor={Matthias C. M. Troffaes, Gero Walter, Dana Kelly},%
pdfkeywords={common-cause failure, alpha-factor model, epistemic uncertainty, conjugate prior, imprecise Dirichlet model}]{hyperref}
\usepackage{mathtools}

\newcommand{\simplex}{\Delta}
\newcommand{\ud}{\,\mathrm{d}}
\newcommand{\lowu}{\underline{u}}
\newcommand{\upu}{\overline{u}}
\newcommand{\lowv}{\underline{v}}
\newcommand{\upv}{\overline{v}}
\newcommand{\lows}{\underline{s}}
\newcommand{\ups}{\overline{s}}
\newcommand{\lowt}{\underline{t}}
\newcommand{\upt}{\overline{t}}
\newcommand{\lowtj}{\lowt_j}
\newcommand{\uptj}{\upt_j}

\newcommand{\pr}{E}
\newcommand{\lpr}{\underline{\pr}}
\newcommand{\upr}{\overline{\pr}}
\newcommand{\GammaFunc}[2]{(#1)_{#2}}

\newcommand{\naturals}{\mathbb{N}_0}

\renewcommand{\theta}{\alpha} %%% switch from stats notation to alpha-factor notation
\renewcommand{\vec}[1]{{\bm#1}}

\title{A Robust Bayesian Approach to Modelling Epistemic Uncertainty in Common-Cause Failure Models}

\author{Matthias C. M. Troffaes}
\address{Durham University, UK}
\email{matthias.troffaes@gmail.com}

\author{Gero Walter}
\address{LMU Munich, Germany}
\email{Gero.Walter@stat.uni-muenchen.de}

\author{Dana Kelly}

\begin{document}

\begin{abstract}
In a standard Bayesian approach to
the alpha-factor model for common-cause failure,
a precise Dirichlet prior distribution
models epistemic uncertainty in the alpha-factors. This Dirichlet prior is
then updated with observed data to obtain a posterior distribution,
which forms the basis for further inferences.

In this paper, we adapt the
%%% no citations in abstract
imprecise Dirichlet model of Walley %\cite{1996:walley::idm}
to represent
epistemic uncertainty in the alpha-factors.  In this approach,
epistemic uncertainty is expressed more cautiously via
lower and upper expectations for each alpha-factor,
along with a learning parameter which determines how quickly
the model learns from observed data.
For this application, we focus on elicitation of the learning parameter, and
find that values in the range of 1 to 10 seem reasonable.
%%% no citations in abstract
The approach is compared with Kelly and Atwood's
%\cite{2011:kelly:atwood}
minimally informative Dirichlet prior for the alpha-factor model,
which incorporated precise
mean values for the alpha-factors, but which was otherwise quite
diffuse.

Next,
we explore the use of a set of Gamma priors
to model epistemic uncertainty in the marginal failure rate,
expressed via a lower and upper expectation for this rate,
again along with a learning parameter.
As zero counts are generally less of an issue here,
we find that the choice of this learning parameter
is less crucial.

Finally, we demonstrate how both epistemic uncertainty models
can be combined to arrive at lower and upper expectations
for all common-cause failure rates.
Thereby,
we effectively provide a full sensitivity analysis
of common-cause failure rates,
properly reflecting
epistemic uncertainty of the analyst
on all levels of the common-cause failure model.
\end{abstract}

\maketitle

\thispagestyle{fancy}

\section{Introduction}

Common-cause failure has been recognized since the time of the Reactor
Safety Study \cite{1975:reactor:safety:study} as a dominant contributor to the
unreliability of redundant systems.  A number of models have been
developed for common-cause failure over the time since the publication
of the Reactor Safety Study, with perhaps the most widely used one
being the Basic Parameter Model, at least in the U.S.~\cite{1988:mosleh::common:cause}.

The alpha-factor parametrisation of this model uses a
multinomial distribution as its aleatory model for observed failures
\cite{1988:mosleh::common:cause}.  The conjugate prior to the multinomial model
is the Dirichlet distribution.  In the standard Bayesian approach, the
analyst specifies the parameters of a precise Dirichlet distribution
to model epistemic uncertainty in the alpha-factors, which are the
parameters of the multinomial aleatory model.  This Dirichlet prior is
then updated with observed data to obtain a precise posterior
distribution, also Dirichlet.

In this paper, we follow \cite{2012:kelly::common:cause:esrel12}, and adapt the
imprecise Dirichlet model of Walley \cite{1996:walley::idm} to represent
epistemic uncertainty in the alpha-factors.  In this approach the
analyst specifies lower or upper expectations (or both) for each
alpha-factor, along with a learning parameter, which determines how quickly
the prior distribution learns from observed data.
We find that values in the range of 1 to 10 seem reasonable for this
application.

Following \cite{2012:kelly::common:cause:esrel12}, 
the approach is compared with that of Kelly and Atwood \cite{2011:kelly:atwood},
which attempted to find a precise Dirichlet prior that was
minimally informative \cite{1996:atwood}, in the sense that it incorporated specified
mean values for the alpha-factors, but which was otherwise quite
diffuse.  The numerical example from \cite{2011:kelly:atwood} is
addressed in the imprecise Dirichlet framework, which can be seen as
an extension of the approach of \cite{2011:kelly:atwood} to the case
where a precise mean for each alpha-factor cannot be specified.

Finally, we address the
problem---not discussed in \cite{2012:kelly::common:cause:esrel12}---of
inference about actual failure rates.
These failure rates are rational functions of the alpha-factors
and the marginal failure rate per component.
Modelling failures as a Poisson process,
we take a Gamma distribution as conjugate prior for the marginal failure rate.
Similar to the procedure for the alpha-factors,
we can model epistemic uncertainty on the marginal failure rate by considering
lower and upper expected prior failure rates,
along with a learning parameter that determines how quickly
the prior distribution learns from observed data.

By combining our epistemic uncertainty models for both the alpha-factors and the marginal failure rate,
we are able to perform a global sensitivity analysis on the common-cause failure rates.
We provide an algorithm that calculates, up to reasonable precision, bounds on these failure rates.
The resulting novel procedure is demonstrated on a simple electrical network reliability problem.

The paper is organized as follows.
Section~\ref{sec:model} reviews the basic parameter model
and its reparametrisation as the alpha-factor model.
Section~\ref{sec:estimation} explores how the parameters of the alpha-factor model
can be estimated, using Dirichlet and Gamma priors.
Section~\ref{sec:epistemic-alpha} discusses the handling of epistemic uncertainty
for the alpha-factors. Two ways to choose a Dirichlet prior
(or sets of Dirichlet priors) starting from epistemic prior expectations of the
alpha-factors are considered.
Throughout, the main ideas are demonstrated on a numerical example.
Section~\ref{sec:epistemic-failure} shows how, similarly to the alpha-factor case,
epistemic uncertainty can be expressed for the marginal failure rate.
A set of conjugate Gamma priors is elicited by considering lower and upper
expected prior marginal failure rates.
Section~\ref{sec:inference} describes an algorithm
that infers bounds on all common-cause failure rates
based on our imprecise alpha-factor model and our imprecise marginal failure rate model.
Section~\ref{sec:example} demonstrates our methodology
on a simple electrical network reliability problem.
Section~\ref{sec:conclusion} ends the paper with some conclusions and
thoughts for further research.

\section{Common-Cause Failure Modelling}
\label{sec:model}

\subsection{The Basic Parameter Model}

Consider a system that consists of $k$ components.
Throughout, we make the following standard assumptions:
(i) repair is immediate, and
(ii) failures follow a Poisson process.

For simplicity, we assume that all $k$ components are exchangeable,
in the sense that they have identical failure rates.
More precisely,
we assume that all events involving \emph{exactly} $j$ components failing
have the same failure rate, which we denote by $q_j$.
This model is called the \emph{basic parameter model},
and we write $\vec{q}$ for $(q_1,\dots,q_k)$.

For example, if we have three components, A, B, and C,
then the rate at which we see only A failing
is equal to the rate at which we see only B failing,
and is also equal to the rate at which we see only C failing;
this failure rate is $q_1$.
Moreover, the rate at which we observe only A and B jointly failing
is equal to the rate at which we observe only B and C jointly failing,
and also equal to the rate at which we observe only A and C jointly failing;
this failure rate is $q_2$.
The rate at which we see all three components jointly failing is $q_3$.

In case of $k$ identical components without common-cause failure modes,
thus each failing independently at rate $\lambda$,
we would have\footnote{This is due to our Poisson assumption, and the assumption of immediate repair: independent Poisson processes never generate events simultaneously when we observe failure times precisely.}
\begin{equation}
  q_1=\lambda\qquad \text{ and }\qquad q_j=0\text{ for }j\ge 2.
\end{equation}
The fact that we allow arbitrary values for the $q_j$ reflects
the lack of independence, and whence,
our modelling of common-cause failures.
At this point, it is worth noting that
we do not actually write down a statistical model
for all possible common-cause failure modes---we
could do so if this information was available,
and in fact, this could render the basic parameter model obsolete,
and allow for more detailed inferences.
In essence, the basic parameter model allows us to statistically model
lack of independence between component failures,
without further detail as to where dependencies arise from:
all failure modes are lumped together, so to speak.

It is useful to note that it is possible, and sometimes necessary,
to relax the exchangeability assumption
to accommodate specific asymmetric cases.
For example, when components are in different state of health,
single failures would clearly not have identical failure rates.
Because the formulas become a lot more complicated,
we stick to the exchangeable case here.

Clearly, to answer typical reliability questions,
such as for instance
``what is the probability that
two or more components fail in the next month?'',
we need $\vec{q}$.
In practice, the following three issues commonly arise. First,
$\vec{q}$ is rarely measured directly,
as failure data is often collected only per component.
Secondly, when direct data about joint failures is available,
typically, this data is sparse,
because events involving more than two components failing simultaneously
are usually quite rare.
Thirdly,
there are usually two distinct sources of failure data,
one usually very large data set related to failure per component,
and one usually much smaller data set
related to joint failures.
For these reasons,
it is sensible to reparametrise the model in terms
of parameters that can be more easily estimated, as follows.

\subsection{The Alpha-Factor Model}

The alpha-factor parametrisation of the basic parameter model
\cite{1988:mosleh::common:cause} starts out with
considering the total failure rate of a component $q_t$,
which could involve failure of any number of components,
that is, this is the rate obtained by looking at just one component,
ignoring everything else.
Clearly,
\begin{equation}\label{eq:qt:in:terms:of:qj}
  q_t=\sum_{j=1}^k\binom{k-1}{j-1}q_j.
\end{equation}
For example, again consider a three component system, A, B, and C.
The rate at which A fails is then
the rate at which only A fails ($q_1$),
plus the rate at which A and B, or A and C fail ($2q_2$),
plus the rate at which all three components fail ($q_3$).

Next, the alpha-factor model introduces
$\theta_j$---the so-called alpha-factor---which
denotes the probability of \emph{exactly} $j$ of the $k$ components
failing given that failure occurs;
in terms of relative frequency,
$\theta_j$ is the fraction of failures
that involve \emph{exactly} $j$ failed components.
We write $\vec{\theta}$ for $(\theta_1,\dots,\theta_k)$.
Clearly,
\begin{equation}\label{eq:alphaj:in:terms:of:qj}
  \alpha_j=\frac{\binom{k}{j}q_j}{\sum_{\ell=1}^k\binom{k}{\ell}q_\ell}.
\end{equation}
For example, again consider A, B, and C.
Then the rate at which exactly one component fails is $3q_1$
(as we have three single components, each of which failing with rate $q_1$),
the rate at which exactly two components fail is $3q_2$
(as we have three combinations of two components,
each combination failing with rate $q_2$),
and the rate at which all components fail is $q_3$.
Translating these rates into fractions,
we arrive precisely at Eq.~\eqref{eq:alphaj:in:terms:of:qj}.

It can be shown that \cite[Table~C-1, p.~C-5]{1988:mosleh::common:cause}:\footnote{Hint: consider $\sum_{j=1}^k j\alpha_j$.}
\begin{equation}\label{eq:qj:in:terms:of:qt}
  q_j=\frac{1}{\binom{k-1}{j-1}}\frac{j\theta_j}{\sum_{\ell=1}^k \ell\theta_\ell}q_t.
\end{equation}
Eqs.~\eqref{eq:qt:in:terms:of:qj}, \eqref{eq:alphaj:in:terms:of:qj},
and~\eqref{eq:qj:in:terms:of:qt}
establish a one-to-one link between the so-called
basic parameter model ($\vec{q}$)
and the alpha-factor model ($q_t$, $\vec{\theta}$).
The benefit of the alpha-factor model over the basic parameter model
lies in its distinction between the total failure rate of a component $q_t$,
for which we generally have a lot of information,
and common-cause failures modelled by $\vec{\theta}$,
for which we generally have very little information.

One of the goals of this paper is to perform a sensitivity analysis,
in the sense of robust Bayes \cite{1984:berger,1990:berger,1991:walley},
over $\vec{\theta}$, and to measure its effects on $q_j$.
Because the $q_j$ are proportional to $q_t$,
in fact, it turns out to take only very little additional effort
to perform a sensitivity analysis
over $\vec{\theta}$ and $q_t$ jointly.
So, although in many cases of practical interest,
we will know $q_t$ quite well, interestingly,
we do not need to assume that we know much at all about $q_t$.

\section{Parameter Estimation}
\label{sec:estimation}

\subsection{Dirichlet Prior for Alpha-Factors}

Suppose that we have observed a sequence of $N$ failure events,
where we have counted the number of components involved
with each failure event,
say $n_j$ of the $N$ observed failure events involved \emph{exactly}
$j$ failed components.
We write $\vec{n}$ for $(n_1,\dots,n_k)$.
In terms of the alpha-factors,
the likelihood for $\vec{n}$ has a very simple form:
\begin{equation}
  \label{eq:multinomial:likelihood}
  \Pr(\vec{n}|\vec{\theta})=\prod_{j=1}^k\theta_j^{n_j},
\end{equation}
which is a multinomial distribution with parameter $\vec{\theta}$.

As mentioned already,
typically, for $j\ge 2$, the $n_j$ are very low, with zero being quite common
for larger $j$.
In such cases, standard techniques such as maximum likelihood for estimating the
alpha-factors fail to produce sensible inferences.
For any inference to be reasonably possible, it has been recognized
\cite{1988:mosleh::common:cause}
that we have to rely on epistemic information, that is, information
which is not just described by the data.

A standard way to include epistemic information in the model is through specification of a Dirichlet
prior for the alpha-factors \cite{1988:mosleh::common:cause}:
\begin{equation}
  \label{eq:dirichlet:prior}
  f(\vec{\theta}|s,\vec{t})\propto\prod_{j=1}^k\theta_j^{st_j-1}
\end{equation}
which is a conjugate prior for the multinomial likelihood specified in Eq.~\eqref{eq:multinomial:likelihood}.
In Eq.~\eqref{eq:dirichlet:prior},
we use Walley's \cite[\S 7.7.3, p.~395]{1991:walley}
$(s,\vec{t})$ notation for the hyperparameters.
Here, $s>0$ and
$\vec{t}\in\simplex$, where $\simplex$ is the $(k-1)$-dimensional unit
simplex:
\begin{equation}
  \simplex=\left\{
  (t_1,\dots,t_k)\colon t_1\ge0,\dots,t_k\ge 0,\,\sum_{j=1}^kt_j=1
  \right\}
\end{equation}
An interpretation for these parameters will be given shortly.
First, let us calculate the posterior density for $\vec{\theta}$:
\begin{equation}
  f(\vec{\theta}|\vec{n},s,\vec{t})
  \propto
  \prod_{j=1}^k\theta_j^{st_j+n_j-1}.
\end{equation}

Of typical interest is for instance the posterior expectation of
the probability $\theta_j$ of
observing $j$ of the $k$ components
failing due to a common cause given that failure occurs:
\begin{equation}
  \label{eq:dirichlet:posterior:predictive}
  E(\theta_j|\vec{n},s,\vec{t})
  =
  \int_{\Delta}\theta_j f(\vec{\theta}|\vec{n},s,\vec{t})\ud\vec{\theta}
  =
  \frac{n_j+st_j}{N+s}
  = \frac{N}{N+s} \frac{n_j}{N} + \frac{s}{N+s} t_j
\end{equation}
where $N=\sum_{j=1}^k n_j$ is the total number of observations.

Eq.~\eqref{eq:dirichlet:posterior:predictive} provides the usual well-known interpretation for the hyperparameters
$s$ and $\vec{t}$:
\begin{itemize}
\item If $N=0$, then $E(\theta_j|s,\vec{t})=t_j$, so $t_j$ is the
  prior expected chance of observing $j$ of the $k$ components
  failing due to a common cause, given that failure occurs.
\item $E(\theta_j|\vec{n},s,\vec{t})$
  is a weighted average of $t_j$ and $n_j/N$ (the proportion of $j$-component failures
  in the $N$ observations), with weights $s$ and $N$, respectively.
  The parameter $s$ thus determines how much data is required for
  the posterior to start moving away from the prior. If $N\ll s$ then
  the prior will weigh more; if $N=s$, then prior and data will weigh
  equally; and if $N\gg s$, then the data will weigh more. In
  particular, $E(\theta_j|\vec{n},s,\vec{t})=t_j$ if $N=0$ (as already
  mentioned), and $E(\theta_j|\vec{n},s,\vec{t})\to\frac{n_j}{N}$ as
  $N\to\infty$.
\end{itemize}

For inference about $q_j$,
which we will discuss in Section~\ref{sec:inference},
we will also need, for natural numbers $p_1$, \dots, $p_k$,
with $P\coloneqq\sum_{j=1}^k p_j$:
\begin{align}
  E\left(\prod_{j=1}^k\theta_j^{p_j}|\vec{n},s,\vec{t}\right)
%  &=
%  \int_{\simplex}\prod_{j=1}^k\theta_j^{p_j} f(\vec{\theta}|\vec{n},s,\vec{t})\ud\vec{\theta}
%  \\
%  &=
%  \frac{\Gamma(N+s)}{\prod_{j=1}^k\Gamma(n_j+st_j)}
%  \int_{\simplex}\prod_{j=1}^k\theta_j^{p_j+n_j+st_j-1} \ud\vec{\theta}
%  \\
%  &=
%  \frac{\Gamma(P+N+s)}{\prod_{j=1}^k\Gamma(p_j+n_j+st_j)}
%  \frac{\prod_{j=1}^k\Gamma(p_j+n_j+st_j)}{\Gamma(P+N+s)}
%  \nonumber\\
%  &\qquad\times
%  \frac{\Gamma(N+s)}{\prod_{j=1}^k\Gamma(n_j+st_j)}
%  \int_{\simplex}\prod_{j=1}^k\theta_j^{p_j+n_j+st_j-1} \ud\vec{\theta}
%  \\
%  &=
%  \frac{\prod_{j=1}^k\Gamma(p_j+n_j+st_j)}{\prod_{j=1}^k\Gamma(n_j+st_j)}
%  \frac{\Gamma(N+s)}{\Gamma(P+N+s)}
%  \\
  &=
  \frac{\prod_{j=1}^k\GammaFunc{n_j+st_j}{p_j}}{\GammaFunc{N+s}{P}}.
  \label{eq:dirichlet:posterior:predictive:2}
\end{align}
where $\GammaFunc{x}{n}$, for $n \in \naturals$, denotes the raising factorial, also known as Pochhammer's symbol \cite[6.1.22, p.~256]{1972:abramowitz}:
%gero to matthias: abramowitz \& stegun have only Pochhammer's symbol as the name
\begin{equation}
  \GammaFunc{x}{n}\coloneqq\frac{\Gamma(x+n)}{\Gamma(x)}=(x+n-1)(x+n-2)\dots (x+1)x.
\end{equation}

By linearity of expectation,
Eq.~\eqref{eq:dirichlet:posterior:predictive:2} allows us to calculate
the expectation of an arbitrary polynomial in $\vec{\theta}$.

\subsection{Per Component Failure Rate}
\label{sec:estimation-failure}

Now we turn to the estimation of $q_t$, the total failure rate per component. 
As mentioned at the start of Section~\ref{sec:model}, %in the introduction,
we assume that failures follow a Poisson process.
Suppose we observe $M$ failures of our component
over a time interval of length $T$.
If $M$ is sufficiently large, then a reasonable
point estimate for $q_t$ would be $M/T$.

Often, that will be enough. However,
in case $M$ is not terribly large,
we can easily propose a conjugate prior for $q_t$.
Specifically,
the likelihood for $M$, given $T$, is:
\begin{equation}\label{eq:poisson:likelihood}
  \Pr(M|q_t,T)
  =
  \frac{(q_t T)^M e^{-q_t T}}{M!}
\end{equation}
which is simply a Poisson distribution with parameter $q_t T$.

A standard way to include epistemic information in the model is through specification of a Gamma
prior \cite{1994:bernardo,2005:quaeghe::expon}:\footnote{We use a non-standard parametrisation to allow easier interpretation of the hyperparameters.}
\begin{equation}
  \label{eq:gamma:prior}
  f(q_t|u,v)\propto q_t^{uv-1} e^{- q_t u},
\end{equation}
which is a conjugate prior for the Poisson likelihood specified in
Eq.~\eqref{eq:poisson:likelihood}.
The posterior density for $q_t$ is:
\begin{equation}
  f(q_t|M,T,u,v)
  \propto
  q_t^{uv+M-1} e^{-q_t(u+T)}
\end{equation}
Of typical interest is the posterior expectation of $q_t$:
\begin{align}
  E(q_t|M,T,u,v)
  &=
  \int_{u,v} f(q_t|M,T,u,v)\ud u\ud v
  =
  \frac{M+uv}{u+T}
  \nonumber
  \\
  \label{eq:gamma:posterior:predictive}
  &=
  \frac{T}{u+T} \frac{M}{T} + \frac{u}{u+T}v
\end{align}

Eq.~\eqref{eq:gamma:posterior:predictive} provides a straightforward interpretation for the hyperparameters
$u$ and $v$, which mimicks our discussion concerning the Dirichlet prior:\footnote{In fact, we arrive at similar interpretations because both priors are members of the canonical exponential family \cite{1994:bernardo,2005:quaeghe::expon}.}
\begin{itemize}
\item If $T=0$, then $E(q_t|u,v)=v$, so $v$ is the
  prior expected failure rate.
\item $E(q_t|M,T,u,v)$
  is a weighted average of $v$ and $M/T$
  (the empirical observed failure rate), with weights $u$ and $T$, respectively.
  The parameter $u$ thus determines for how long we need to observe the process %any data
  %before the posterior to start moving away from the prior.
  until the posterior starts to move away from the prior.
  If $T\ll u$ then the prior will weigh more; if $T=u$, then prior and data will weigh
  equally; and if $T\gg u$, then the data will weigh more. In
  particular, $E(q_t|M,T,u,v)=v$ if $T=0$ (as already
  mentioned), and $E(q_t|M,T,u,v)\to\frac{M}{T}$ as
  $T\to\infty$.
\end{itemize}

\section{Handling Epistemic Uncertainty in Alpha-Factors}
\label{sec:epistemic-alpha}

Crucial to reliable inference in the alpha-factor model is proper
modelling of epistemic uncertainty about failures, which is in the
above approach expressed through the $(s,\vec{t})$ parameters.
We focus on two methods for elicitation of these parameters,
and the inferences that result from them.

Throughout, we will use the following example, which is taken
from Kelly and Atwood \cite{2011:kelly:atwood}.
Consider a system with four redundant components ($k=4$).
The probability of
$j$ out of $k$ failures, given that failure has happend,
was denoted by $\theta_j$.
We assume that the analyst's prior expectation $\mu_{\text{spec},j}$
for each $\theta_j$ is:
\begin{align}
  \label{eq:example:muspec}
  \mu_{\text{spec},1}&=0.950
  &
  \mu_{\text{spec},2}&=0.030
  &
  \mu_{\text{spec},3}&=0.015
  &
  \mu_{\text{spec},4}&=0.005
\end{align}
We have 36 observations, in which 35 showed one component failing,
and 1 showed two components failing:
\begin{align*}
  n_1&=35
  &
  n_2&=1
  &
  n_3&=0
  &
  n_4&=0
\end{align*}

\subsection{Constrained Non-Informative Prior}

Atwood \cite{1996:atwood} studied priors for the binomial model which
maximise entropy (and whence, are `non-informative')
whilst constraining the mean to a specific value.
Although these priors are not conjugate, Atwood \cite{1996:atwood} showed that they
can be well approximated by Beta distributions, which are conjugate.
Kelly and Atwood \cite{2011:kelly:atwood} applied this approach
to the multinomal model with conjugate Dirichlet priors,
by choosing a constrained non-informative prior for the marginals
of the Dirichlet---which are Beta.
This leads to an over-specified system of equalities, which can
be solved via least-squares optimisation.

For the problem we are interested in, $\mu_{\text{spec},1}$ is close to $1$.
In this case, the solution of the least-squares problem
turns out to be close to:
\begin{equation}
 \label{eq:constrainednoninformative}
 \begin{split}
  t_j&=\mu_{\text{spec},j} \text{ for all $j\in\{1,\dots,k\}$} \\
  s&=\frac{1}{2(1-\mu_{\text{spec},1})}
 \end{split}
\end{equation}
For our example, this means that $s=10$
\cite[p.~400, \S 3]{2011:kelly:atwood}. An obvious calculation reveals
that, under this prior \cite[p.~401, \S 3.1]{2011:kelly:atwood}:
\begin{align*}
  E(\theta_1|\vec{n},s,\vec{t})&=\frac{35+9.5}{36+10}=0.967
  &
  E(\theta_2|\vec{n},s,\vec{t})&=\frac{1+0.3}{36+10}=0.028
  \\
  E(\theta_3|\vec{n},s,\vec{t})&=\frac{0+0.15}{36+10}=0.003
  &
  E(\theta_4|\vec{n},s,\vec{t})&=\frac{0+0.05}{36+10}=0.001
\end{align*}
Kelly and Atwood
\cite[p.~402, \S 4]{2011:kelly:atwood}
compare these results against a large number of other choices of priors,
and note that the posterior resulting from Eq.~\eqref{eq:constrainednoninformative} seems too strongly influenced
by the prior, particularly in the presence of zero counts.
For instance, the uniform prior is a Dirichlet distribution with hyperparameters
$t_j=0.25$ and $s=4$, which gives:
\begin{align*}
  E(\theta_1|\vec{n},s,\vec{t})&=\frac{35+1}{36+4}=0.9
  &
  E(\theta_2|\vec{n},s,\vec{t})&=\frac{1+1}{36+4}=0.05
  \\
  E(\theta_3|\vec{n},s,\vec{t})&=\frac{0+1}{36+4}=0.025
  &
  E(\theta_4|\vec{n},s,\vec{t})&=\frac{0+1}{36+4}=0.025
\end{align*}
Jeffrey's prior is again a Dirichlet distribution with hyperparameters
$t_j=0.125$ and $s=4$, which gives:
\begin{align*}
  E(\theta_1|\vec{n},s,\vec{t})&=\frac{35+0.5}{36+4}=0.8875
  &
  E(\theta_2|\vec{n},s,\vec{t})&=\frac{1+0.5}{36+4}=0.0375
  \\
  E(\theta_3|\vec{n},s,\vec{t})&=\frac{0+0.5}{36+4}=0.0125
  &
  E(\theta_4|\vec{n},s,\vec{t})&=\frac{0+0.5}{36+4}=0.0125
\end{align*}

The degree of variation in the posterior under different priors
is evidently somewhat alarming.
In the next section, we aim to robustify the model
by using sets of priors from the start.

\subsection{Imprecise Dirichlet Model}
\label{sec:idm}

\subsubsection{Near-Ignorance Model}
\label{sec:idm-nearignorance}

In case no prior information is available, Walley proposes as a so-called
\emph{near-ignorance prior} a set of Dirichlet priors, with hyperparameters constrained
to the set:
\begin{equation*}
  \mathcal{H}=\{(s,\vec{t})\colon\vec{t}\in\simplex\}
\end{equation*}
for some fixed value of $s$, which determines the learning speed of
the model \cite[p.~218, \S 5.3.2]{1991:walley} \cite[p.~9, \S
  2.3]{1996:walley::idm}.

\subsubsection{General Model}
\label{sec:idm-general}

When prior information is available, more generally, we may assume
that we can specify a subset $\mathcal{H}$ of
$(0,+\infty)\times\Delta$.
Following Walley's suggestions
\cite[p.~224, \S 5.4.3]{1991:walley} \cite[p.~32, \S 6]{1996:walley::idm},
we take
\begin{equation}
  \label{eq:hyperparams:boxmodel}
  \mathcal{H}
  =
  \left\{
  (s,\vec{t})
  \colon
  s\in[\lows,\ups],\,
  \vec{t}\in\Delta,\,
  t_j\in[\lowtj,\uptj]
  \right\}
\end{equation}
where the analyst has to specify the bounds
$[\lowtj,\uptj]$ for each $j\in\{1,\dots,k\}$,
and $[\lows,\ups]$.

The posterior lower and upper expectations of $\theta_j$ are:
\begin{align}
  \label{eq:idm-update-lower}
  \lpr(\theta_j|\vec{n},\mathcal{H})
  &=
  \min\left\{
    \frac{n_j+\lows\lowtj}{N+\lows},
    \frac{n_j+\ups\lowtj}{N+\ups}
  \right\}
  =
  \begin{cases}
    \frac{n_j+\lows\lowtj}{N+\lows} & \text{if }\lowtj\ge n_j/N \\[1ex]
    \frac{n_j+\ups\lowtj}{N+\ups} & \text{if }\lowtj\le n_j/N
  \end{cases}
  \\
  \label{eq:idm-update-upper}
  \upr(\theta_j|\vec{n},\mathcal{H})
  &=
  \max\left\{
    \frac{n_j+\lows\uptj}{N+\lows},
    \frac{n_j+\ups\uptj}{N+\ups}
  \right\}
  =
  \begin{cases}
    \frac{n_j+\ups\uptj}{N+\ups} & \text{if }\uptj\ge n_j/N \\[1ex]
    \frac{n_j+\lows\uptj}{N+\lows} & \text{if }\uptj\le n_j/N
  \end{cases}
\end{align}

For the model to be of any use, we must be able to elicit the bounds.
The interval $[\lowtj,\uptj]$ simply represents bounds on the prior
expectation of the chance $\theta_j$.

\paragraph{Fixed Learning Parameter}

Typically, the learning parameter $s$ is taken to be $2$
(not without controversy; see
insightful discussions in \cite{1996:walley::idm}).
One might therefore be tempted to using
the same prior expectations $t_j$ for the $\theta_j$ as above (Eq.~\eqref{eq:example:muspec}), with $s=2$,
resulting in the following posterior expectations:
\begin{align*}
  E(\theta_1|\vec{n},s,\vec{t})&=\frac{35+1.9}{36+2}=0.971
  &
  E(\theta_2|\vec{n},s,\vec{t})&=\frac{1+0.06}{36+2}=0.028
  \\
  E(\theta_3|\vec{n},s,\vec{t})&=\frac{0+0.03}{36+2}=0.0007
  &
  E(\theta_4|\vec{n},s,\vec{t})&=\frac{0+0.01}{36+2}=0.0002
\end{align*}
Whence, for this example, it is obvious that $s=2$ is an excessively
poor choice: the posterior expectations in case of zero counts are
pulled way too much towards zero. One might suspect that this is partly due to the
strong prior information, that is, the knowledge of $t_j$. However,
even if we interpret the given probabilities as bounds, say:
\begin{subequations}
  \label{eq:example:tintervals}
\begin{align} %%% same layout as below
  [\lowt_1,\upt_1]&=[0.950,1]
  \\
  [\lowt_2,\upt_2]&=[0,0.030]
  \\
  [\lowt_3,\upt_3]&=[0,0.015]
  \\
  [\lowt_4,\upt_4]&=[0,0.005]
\end{align}
\end{subequations}
we still find:
\begin{subequations}
\begin{align}
  [
  \lpr(\theta_1|\vec{n},\mathcal{H})%=\frac{35+1.9}{36+2}
  ,
  \upr(\theta_1|\vec{n},\mathcal{H})%=\frac{35+2}{36+2}
  ]
  &=
  [
  0.971
  ,
  0.974
  ]
  \\
  [
  \lpr(\theta_2|\vec{n},\mathcal{H})%=\frac{1}{36+2}
  ,
  \upr(\theta_2|\vec{n},\mathcal{H})%=\frac{1+0.06}{36+2}
  ]
  &=
  [
  0.026
  ,
  0.028
  ]
  \\
  [
  \lpr(\theta_3|\vec{n},\mathcal{H})%=\frac{0}{36+2}
  ,
  \upr(\theta_3|\vec{n},\mathcal{H})%=\frac{0+0.03}{36+2}
  ]
  &=
  [
  0
  ,
  0.0007
  ]
  \\
  [
  \lpr(\theta_4|\vec{n},\mathcal{H})%=\frac{0}{36+2}
  ,
  \upr(\theta_4|\vec{n},\mathcal{H})%=\frac{0+0.01}{36+2}
  ]
  &=
  [
  0
  ,
  0.0002
  ]
\end{align}
\end{subequations}
Clearly, only the posterior inferences about $\theta_1$
(and perhaps also $\theta_2$) seem reasonable.
We conclude that \emph{the imprecise Dirichlet model with $s=2$ learns
  too fast from the data in case of zero counts}.

On the one hand, when counts are sufficiently far from zero, the
posterior probability with $s=2$, and perhaps even $s=1$ or $s=0$,
seem appropriate. For
zero counts, however, a larger value of $s$ seems mandatory. Therefore,
it seems logical to pick an interval for $s$.

A further argument for choosing an interval for $s$, in case of an
informative set of priors, is provided by Walley \cite[p.~225, \S
5.4.4]{1991:walley}: a larger value of $\ups$ ensures that the
posterior does not move away too fast from the prior, which is
particularly important for zero counts, and the difference between
$\lows$ and $\ups$ effectively results in greater posterior
imprecision if $n_j/N \notin [\lowtj,\uptj]$. %in case of prior-data conflict.

To see this, note that, if $\lowtj \le n_j/N \le \uptj$, it follows from Eqs. \eqref{eq:idm-update-lower} and \eqref{eq:idm-update-upper} that both lower and upper posterior
expectation are calculated using $\ups$. When $n_j/N \le \lowtj$ (or
$\uptj \le n_j/N$), the lower (upper) posterior expectation is calculated
using $\lows$ instead, which is nearer to $n_j/N$ due to the lower weight $\lows$ for the prior bound $\lowtj$ ($\uptj$).
%resulting in a wider posterior expectation interval.
The increased imprecision reflects the conflict between the prior assignment
$[\lowtj,\uptj]$ and the observed fraction $n_j/N$, and this is referred to as
\emph{prior-data conflict} (also see \cite{2009:walter}).

\paragraph{Interval for Learning Parameter}

We follow Good \cite[p.~19]{1965:good}
(as suggested by Walley \cite[Note~5.4.1, p.~524]{1991:walley}),
and reason about posterior expectations of hypothetical data
to elicit $\lows$ and $\ups$;
also see \cite[p.~219, \S 5.3.3]{1991:walley} for further
discussion on elicitation on $s$---our approach is similar, but
simpler for the case under study. We assume that $\upt_1=1$ and
$\lowtj=0$ for all $j\ge 2$.

The upper probability of multiple ($j\ge 2$) failed components in trial
$m+1$, given one ($j=1$) failed component in all of the first $m$ trials, is
\begin{equation*}
  \upr(\theta_j|n_1=m,N=m,\mathcal{H})=\frac{\ups\uptj}{m+\ups}
\end{equation*}
(Note: there is no prior-data conflict in this case.)
Whence, for the above probability to reduce to $\uptj/2$ (i.e., to
reduce the prior upper probability by half), we need that $m=\ups$.
In other words, \emph{$\ups$ is the number of one-component failures required to
  reduce the upper probabilities of multi-components failure by half}.

Conversely, the lower probability of one ($j=1$) failed component
in trial $m+1$, given only multiple ($j\ge 2$) failed components in the first $m$ trials,
is
\begin{equation*}
  \lpr(\theta_1|n_1=0,N=m,\mathcal{H})=\frac{\lows\lowt_1}{m+\lows}
\end{equation*}
(Note: there is strong prior-data conflict in this case.)
In other words, \emph{$\lows$  is the number of multi-component failures required to
  reduce the lower probability of one-component failure by half}.
Note that, in this case, a few alternative interpretations present
themselves. First, for $j \ge 2$,
\begin{equation*}
  \upr(\theta_j|n_j=m,N=m,\mathcal{H})=\frac{m+\lows\uptj}{m+\lows}
\end{equation*}
In other words, \emph{$\lows$ is also the number of $j$-component failures
  required to increase the upper probability of $j$ components failing
  to $(1+\uptj)/2$} (generally, this will be close to $1/2$,
provided that $\uptj$ is close to zero).
Secondly, for $j \ge 2$,
\begin{equation*}
  \lpr(\theta_j|n_j=m,N=m,\mathcal{H})=\frac{m}{m+\ups}
\end{equation*}
so \emph{$\ups$ is also the number of multi-component failures required to
  increase the lower probability of multi-component failures to a half}.

Any of these counts seem well suited for elicitation, and are easy to
interpret. As a guideline, we suggest the following easily remembered
rules:
\begin{itemize}
\item $\ups$ is the number of one-component failures required to reduce the
  upper probabilities of multi-component failures by
  half, and
\item $\lows$ is the number of multi-component failures required to reduce the lower
  probability of one-component failures by half.
\end{itemize}
Taking the above interpretation, the difference between $\ups$ and
$\lows$ reflects the fact that the rate at which we reduce upper
probabilities is less than the rate at which we reduce lower
probabilities, and thus reflects a level of caution in our model.

Coming back to our example, reasonable values are $\lows=1$ (if we
immediately observe multi-component failures, we might be quite keen
to reduce our lower probability for one-component failure) and
$\ups=10$ (we are happy to halve our upper probabilities of
multi-component failures after observing $10$ one-component
failures). With these values,
when taking for $t_j$ the values given in Eq.~\eqref{eq:example:muspec},
we find the following posterior lower
and upper expectations of $\theta_j$:
\begin{subequations}
\begin{align}
  [
  \lpr(\theta_1|\vec{n},\mathcal{H})%=\frac{35+9.5}{36+10}
  ,
  \upr(\theta_1|\vec{n},\mathcal{H})%=\frac{35+0.95}{36+1}
  ]
  &=
  [
  0.967
  ,
  0.972
  ]
  \\
  [
  \lpr(\theta_2|\vec{n},\mathcal{H})%=\frac{1+0.03}{36+1}
  ,
  \upr(\theta_2|\vec{n},\mathcal{H})%=\frac{1+0.3}{36+10}
  ]
  &=
  [
  0.0278
  ,
  0.0283
  ]
  \\
  [
  \lpr(\theta_3|\vec{n},\mathcal{H})%=\frac{0+0.015}{36+1}
  ,
  \upr(\theta_3|\vec{n},\mathcal{H})%=\frac{0+0.15}{36+10}
  ]
  &=
  [
  0.00041
  ,
  0.00326
  ]
  \\
  [
  \lpr(\theta_4|\vec{n},\mathcal{H})%=\frac{0+0.005}{36+1}
  ,
  \upr(\theta_4|\vec{n},\mathcal{H})%=\frac{0+0.05}{36+10}
  ]
  &=
  [
  0.00014
  ,
  0.00109
  ]
\end{align}
\end{subequations}
These bounds indeed reflect caution in inferences
where zero counts have occurred ($j=3$ and $j=4$), with upper
expectations considerably larger as compared to the model with fixed $s$,
while still giving a reasonable expectation interval for the probability of
one-component failure.

If we desire to specify our initial bounds for $t_j$ more conservatively,
as in Eqs.~\eqref{eq:example:tintervals}, we find similar results:
\begin{subequations}\label{eq:alphabounds}
\begin{align}
  [
  \lpr(\theta_1|\vec{n},\mathcal{H})%=\frac{35+9.5}{36+10}
  ,
  \upr(\theta_1|\vec{n},\mathcal{H})%=\frac{35+10}{36+10}
  ]
  &=
  [
  0.967
  ,
  0.978
  ]
  \\
  [
  \lpr(\theta_2|\vec{n},\mathcal{H})%=\frac{1+0}{36+1}
  ,
  \upr(\theta_2|\vec{n},\mathcal{H})%=\frac{1+0.3}{36+10}
  ]
  &=
  [
  0.0270
  ,
  0.0283
  ]
  \\
  [
  \lpr(\theta_3|\vec{n},\mathcal{H})%=\frac{0+0}{36+1}
  ,
  \upr(\theta_3|\vec{n},\mathcal{H})%=\frac{0+0.15}{36+10}
  ]
  &=
  [
  0
  ,
  0.00326
  ]
  \\
  [
  \lpr(\theta_4|\vec{n},\mathcal{H})%=\frac{0+0}{36+1}
  ,
  \upr(\theta_4|\vec{n},\mathcal{H})%=\frac{0+0.05}{36+10}
  ]
  &=
  [
  0
  ,
  0.00109
  ]
\end{align}
\end{subequations}

\section{Handling Epistemic Uncertainty in Marginal Failure Rate}
\label{sec:epistemic-failure}

Before we can consider inferences on the common-cause failure rates $q_j$,
we will briefly explain how we express epistemic uncertainty on the marginal failure rate $q_t$.
As seen in Section~\ref{sec:estimation-failure}, we will use conjugate Gamma priors
with hyperparameters $u$ and $v$, where $v$ is the prior failure rate parameter,
and $u$ determines the learning speed.
Similarly to the alpha-factor case, we can express vague prior information on $q_t$
by considering sets of priors, which are generated by sets of hyperparameters,
i.e., we specify a parameter set $\mathcal{J} \subseteq (0, \infty) \times (0, \infty)$.
Unlike Section~\ref{sec:idm-nearignorance},
here $\mathcal{J} = \{u\}\times (0, \infty)$,
for some fixed value of $u$,
does not lead to a practically useful near-ignorant set of priors,
as then $\upr(q_t|M, T, \mathcal{J}) = \infty$ for any $M$ and $T$.
In practice, it should not be a big issue to find bounds $[\lowv, \upv]$ for the prior expected marginal failure rate.

Similarly to Eqs.~\eqref{eq:idm-update-lower} and \eqref{eq:idm-update-upper},
when $\mathcal{J}=[\lowu,\upu]\times[\lowv,\upv]$,
the posterior lower and upper expectations of $q_t$ are 
\begin{align}
  \label{eq:igm-update-lower}
  \lpr(q_t|M, T, \mathcal{J})
  &=
  \min\left\{
    \frac{M+\lowu\lowv}{T+\lowu},
    \frac{M+\upu\lowv}{T+\upu}
  \right\}
  =
  \begin{cases}
    \frac{M+\lowu\lowv}{T+\lowu} & \text{if }\lowv\ge M/T \\[1ex]
    \frac{M+\upu\lowv}{T+\upu} & \text{if }\lowv\le M/T
  \end{cases}
  \\
  \label{eq:igm-update-upper}
  \upr(q_t|M, T, \mathcal{J})
  &=
  \max\left\{
    \frac{M+\lowu\upv}{T+\lowu},
    \frac{M+\upu\upv}{T+\upu}
  \right\}
  =
  \begin{cases}
    \frac{M+\upu\upv}{T+\upu} & \text{if }\upv\ge M/T \\[1ex]
    \frac{M+\lowu\upv}{T+\lowv} & \text{if }\upv\le M/T
  \end{cases}
\end{align}

To elicit bounds for the learning parameter $u$, 
similar considerations as in Section~\ref{sec:idm-general} can be made.
Assuming $\lowv = 0$, the posterior lower expectation for $q_t$ is
\begin{equation}\label{eq:elicit-upv}
  \lpr(q_t|M,T, \mathcal{J})=\frac{M}{T+\upu}
\end{equation}
(Note: there is no prior-data conflict in this case.)
Whence, $\upu$ is the amount of time needed to observe the process until we raise
the lower expectation of $q_t$ from $0$ to half of the observed failure rate $M/T$.

Conversely, assuming $\lowv > 0$, and no failures at all during time $T$,
the posterior lower expectation for $q_t$ is 
\begin{equation}\label{eq:elicit-lowv}
  \lpr(q_t|M = 0, T, \mathcal{J}) = \frac{\lowu\lowv}{T+\lowu}
  =
  \frac{\lowv}{\frac{T}{\lowu}+1}
\end{equation}
(Note: prior-data conflict is present in this case.)
Whence, $\lowu$ is the time needed to observe the process---without any failures---until
$\lowv$ is reduced by half.

Contrary to the situation in Section~\ref{sec:epistemic-alpha},
zero counts are much less of a concern
when estimating the marginal failure rate.
Whence, for sake of simplicity,
it might therefore suffice to consider parameter sets of the form
\begin{equation}
\label{eq:gammaprior:box}
\mathcal{J} = \{u\} \times [\lowv, \upv]
\end{equation}
only.
Both Eqs.~\eqref{eq:elicit-upv} and~\eqref{eq:elicit-lowv}
can then serve to determine $u=\lowu=\upu$.

A numerical example will be given in Section~\ref{sec:example}.

\section{Inference on Failure Rates}
\label{sec:inference}

\subsection{Expected Failure Rates}
\label{sec:inference:expectation}

For inference on the failure rates $q_j$, we will now combine our models
for alpha-factors and marginal failure rate by using Eq.~\eqref{eq:qj:in:terms:of:qt}. 
The problem in doing this is that there is, as far as we know,
no immediate closed expression for the posterior expectation of $q_j$,
because Eq.~\eqref{eq:qj:in:terms:of:qt}
is a rational function of $\vec{\theta}$.
However, naively,
we can approximate it using Taylor expansion.
Specifically,
\begin{align}
  q_j
  &=\frac{1}{\binom{k-1}{j-1}}\frac{j\theta_j}{\sum_{\ell=1}^k \ell\theta_\ell}q_t
  \\
  &=\frac{1}{\binom{k-1}{j-1}}\frac{j\theta_j}{\sum_{\ell=1}^k (\theta_\ell + (\ell-1)\theta_\ell)}q_t
  \\
  &=\frac{1}{\binom{k-1}{j-1}}\frac{j\theta_j}{1+\sum_{\ell=2}^k (\ell-1)\theta_\ell}q_t
  \\
  \intertext{and, as long as $\sum_{\ell=2}^k (\ell-1)\theta_\ell<1$---this is always true if $k\le 2$; for larger $k$, it is usually true because $\theta_\ell$ is usually very small for $\ell\ge 3$---we can
  use the Taylor expansion $1/(1+x)=1-x+x^2-x^3+\dots$ (valid for $|x|<1$), to arrive at:}
  &=
  \frac{1}{\binom{k-1}{j-1}}j\theta_j
  \left[
    1
    -\sum_{\ell=2}^k (\ell-1)\theta_\ell
    +\left\{\sum_{\ell=2}^k (\ell-1)\theta_\ell\right\}^2
    -\dots
  \right]
  q_t
  \label{eq:qj:taylor:expansion}
\end{align}
The posterior expectation of Eq.~\eqref{eq:qj:taylor:expansion}
can now be evaluated,
using Eqs.~\eqref{eq:dirichlet:posterior:predictive:2}
and~\eqref{eq:gamma:posterior:predictive},
under the usual assumption that $q_t$ is independent of the alpha-factors.
%%% note: no clear reference for this, but seems to be assumed

\begin{table}
  \begin{center}
    \begin{tabular}{c|cccc}
      $x$ & $\frac{1}{1+x}$ & $1-x$ & $1-x+x^2$ \\
      \hline
      0.0 & 1.0  & 1.0 & 1.0  \\
      0.1 & 0.91 & 0.9 & 0.91 \\
      0.2 & 0.83 & 0.8 & 0.84 \\
      0.3 & 0.77 & 0.7 & 0.79 \\
      0.4 & 0.71 & 0.6 & 0.76 \\
      0.5 & 0.67 & 0.5 & 0.75 \\
      0.6 & 0.63 & 0.4 & 0.76 \\
      0.7 & 0.59 & 0.3 & 0.79 \\
      0.8 & 0.56 & 0.2 & 0.84 \\
      0.9 & 0.53 & 0.1 & 0.91
    \end{tabular}
    \caption{Accuracy of first and second order Taylor approximations.}
    \label{tab:taylor}
  \end{center}
\end{table}
To get a better idea of accuracy,
Table~\ref{tab:taylor} tabulates first and second order approximations.
For example, second order approximation remains fairly accurate
for $\sum_{\ell=2}^k (\ell-1)\theta_\ell<0.5$,
and first order approximation
for $\sum_{\ell=2}^k (\ell-1)\theta_\ell<0.3$.

An obvious issue with Taylor approximation is that
the domain of integration includes values for $\vec{\alpha}$
where the Taylor series does not converge. However,
it is easy to see that, for any $x\ge 0$ (not just those for which $|x|<1$):
\begin{align}
  0&\le (1-x+x^2-\dots+(-x)^{p})-\frac{1}{1+x}\le x^{p+1}
  &&\text{for even $p$, and}
  \\
  0&\le \frac{1}{1+x}-(1-x+x^2-\dots+(-x)^{p})\le x^{p+1}
  &&\text{for odd $p$}.
\end{align}
Therefore, for any non-negative random variables $x$ and $y$:
\begin{align}
  0\le
  & E[y(1-x+x^2-\dots+(-x)^{p})]
  - E\left(\frac{y}{1+x}\right)
  \le E(yx^{p+1})
  &&\text{for even $p$, and}
  \\
  0\le & E\left(\frac{y}{1+x}\right) - E[y(1-x+x^2-\dots+(-x)^{p})]
  \le E(yx^{p+1})
  &&\text{for odd $p$}.
\end{align}
So as long as the expectation of $yx^{p+1}$ is small enough,
taking the expectation over the Taylor expansion, of order $p$,
will provide a reasonable approximation.

As an example, for the special but important case of $k=2$,
we derive expressions for the posterior expectation of $q_1$ and $q_2$,
under second order approximation,
along with error term:
\begin{align}
  &E(q_1|\vec{n},s,\vec{t};M,T,u,v)
  \\
  &\approx
  E\left(
  \theta_1\left[1-\theta_2+\theta_2^2\right]
  \middle|
  \vec{n},s,\vec{t}
  \right)
  E(q_t|M,T,u,v)
  \\
  &=
  \left(
    \frac{n_1+st_1}{N+s}
    -
    \frac{\GammaFunc{n_1+st_1}{1}\GammaFunc{n_2+st_2}{1}}{\GammaFunc{N+s}{2}}
    +
    \frac{\GammaFunc{n_1+st_1}{1}\GammaFunc{n_2+st_2}{2}}{\GammaFunc{N+s}{3}}
  \right)
  \frac{M+uv}{u+T}
  \\
  &=
  \frac{n_1+st_1}{N+s}
  \left(
    1
    -
    \frac{n_2+st_2}{N+s+1}\left(
      1
      -
      \frac{n_2+st_2+1}{N+s+2}
    \right)
  \right)
  \frac{M+uv}{u+T}
\end{align}
%%% matthias to gero: in the above form, it is easy to see how it extends to arbitrary orders of approximation, also it's the most efficient way to calculate
%%% gero to matthias: yes, the intermediate step with the pochhammer symbol clarifies a lot. If possible regarding page limits, we should keep it in the final version.
up to an absolute expected error less than
\begin{multline}
  E\left(
  \theta_1\theta_2^3
  \middle|
  \vec{n},s,\vec{t}
  \right)
  E(q_t|M,T,u,v)
  \\
  =
  \frac{(n_1+st_1)(n_2+st_2)(n_2+st_2+1)(n_2+st_2+2)}{%
    (N+s)(N+s+1)(N+s+2)(N+s+3)}
  \frac{M+uv}{u+T}
\end{multline}
and similarly,
\begin{align}
  &E(q_2|\vec{n},s,\vec{t};M,T,u,v)
  \\
  &\approx
  E\left(
  2\theta_2\left[1-\theta_2+\theta_2^2\right]
  \middle|
  \vec{n},s,\vec{t}
  \right)
  E(q_t|M,T,u,v)
  \\
  &=
  2
  \left(
  \frac{n_2+st_2}{N+s}
  -
  \frac{\GammaFunc{n_2+st_2}{2}}{\GammaFunc{N+s}{2}}
  +
  \frac{\GammaFunc{n_2+st_2}{3}}{\GammaFunc{N+s}{3}}
  \right)
  \frac{M+uv}{u+T}
  \\
  &=
  2\frac{n_2+st_2}{N+s}
  \left(
    1
    -
    \frac{n_2+st_2+1}{N+s+1}
    \left(
      1-\frac{n_2+st_2+2}{N+s+2}
    \right)
  \right)
  \frac{M+uv}{u+T}
\end{align}
up to an absolute expected error less than
\begin{multline}
  E\left(
  2\theta_2\theta_2^3
  \middle|
  \vec{n},s,\vec{t}
  \right)
  E(q_t|M,T,u,v)
  \\
  =
  2\frac{(n_2+st_2)(n_2+st_2+1)(n_2+st_2+2)(n_2+st_2+3)}{%
    (N+s)(N+s+1)(N+s+2)(N+s+3)}
  \frac{M+uv}{u+T}
\end{multline}

\subsection{Sensitivity Analysis}
\label{sec:inference:sensitivity}

As mentioned in Sections~\ref{sec:epistemic-alpha} and \ref{sec:epistemic-failure},
due to epistemic uncertainty, generally,
an analyst specifies bounds for the hyperparameters
$s$, $\vec{t}$, $u$, and $v$.
The parameter sets are, as before, denoted by $\mathcal{H}$ and $\mathcal{J}$.
As we assumed $q_t$ and $\vec{\theta}$ to be independent (see Section~\ref{sec:inference:expectation}),
we can seperate the analysis into two simpler problems.
We first calculate lower and upper bounds on the expectation of the terms depending on $\theta$,
based on the results from Sections~\ref{sec:epistemic-alpha} and~\ref{sec:inference:expectation}.
Independently, we calculate lower and upper bounds on the expectation of $q_t$
as we did in Section~\ref{sec:epistemic-failure}.
These bounds uniquely determine $\lpr(q_j|\vec{n},M,T,\mathcal{H},\mathcal{J})$
and
$\upr(q_j|\vec{n},M,T,\mathcal{H},\mathcal{J})$, as follows.

For convience of notation, define
\begin{equation}\label{eq:g}
  g_j(\vec{\theta})
  \coloneqq
  \frac{1}{\binom{k-1}{j-1}}\frac{j\theta_j}{\sum_{\ell=1}^k \ell\theta_\ell}.
\end{equation}
Clearly, by Eq.~\eqref{eq:qj:in:terms:of:qt},
\begin{equation}
  q_j=g_j(\vec{\theta})q_t
\end{equation}
so,
\begin{align}
  \lpr(q_j|\vec{n},M,T,\mathcal{H},\mathcal{J})
  =
  \lpr(g_j(\vec{\theta})|\vec{n},\mathcal{H})
  \lpr(q_t|M,T,\mathcal{J})
\end{align}
where
\begin{align}
  \label{eq:optim-gjalpha}
  \lpr(g_j(\vec{\theta})|\vec{n},\mathcal{H})
  &=
  \min_{(s,\vec{t})\in\mathcal{H}}
  E(g_j(\vec{\theta})|\vec{n},s,\vec{t})
  \\
  \label{eq:optim-qt}
  \lpr(q_t|M,T,\mathcal{J})  
  &=
  \min_{(u,v)\in\mathcal{J}}
  E(q_t|M,T,u,v)
\end{align}
Similar expressions for the upper expectation hold as
well by simply replacing $\min$ by $\max$ at all instances.
When using Taylor approximation,
an upper bound on the error term follows readily as well
(see example in Section~\ref{sec:example}).

As seen in Section~\ref{sec:epistemic-failure},
the optimisation problem in Eq.~\eqref{eq:optim-qt} %over $\mathcal{J}$ 
(and its counterpart of the upper bound)
can be done exactly, using
Eqs.~\eqref{eq:igm-update-lower} and~\eqref{eq:igm-update-upper}.
In contrast,
the optimisation problem in Eq.~\eqref{eq:optim-gjalpha} %over $\mathcal{H}
(and its counterpart of the upper bound)
is not so obvious,
and we have to rely on standard numerical algorithms
for non-linear optimisation.
However, the particular form of $\mathcal{H}$ we assumed
in Section~\ref{sec:idm-general} (see Eq.~\eqref{eq:hyperparams:boxmodel})
makes this optimisation problem fairly easily solvable by computer.

\section{Example}
\label{sec:example}

To conclude the paper,
we demonstrate our methodology on a simple
electrical network reliability problem.
Numbers are fictional,
yet are representative of a typical network
in the North-East of England.

A group of customers is supplied with power
from two identical distribution lines.
Supply is lost when both lines fail.
Nationwide statistics show typical failure rate of
similar distribution lines to be within $\pm$50\% of $0.35$ per year.
Nationwide statistics also show the typical fraction of double failures
to be between $10\%$ and $20\%$.
On the actual system under study,
over the last 12 years,
11 failures were observed, 3 of which were double failures.

A typical quantity one would be interested in is $q_2=g_2(\vec{\theta})q_t$,
the rate of double failures,
as this is also the rate at which customers lose power.

For the lower and upper expectation of $q_t$,
we take $u=3$ (in years)---this means that we need about 3 years of data before we start moving away from our prior.
For $v$, we take $[0.175,0.525]$, that is all values within $\pm$50\%
of the nationwide average $0.35$.
We find:\footnote{We have two lines, each observed 12 years, with 8 single failures on either line, and 3 failures occurring in both lines; whence, marginally, we observed $8+3\times 2=14$ failures of a distribution line  over a total timespan of 24 years.}
\begin{align}
  \lpr(q_t|M,T,\mathcal{J})&=\frac{14+3\times 0.175}{24+3}=0.538
  \\
  \upr(q_t|M,T,\mathcal{J})&=\frac{14+3\times 0.525}{24+3}=0.577
\end{align}

For the lower and upper expectation of $g_2(\vec{\theta})$,
we take $[1,4]$ for $s$,\footnote{In this simple example, we have no zero counts, so we can do with a lower upper bound for $s$.}
$[0.8,0.9]$ for $t_1$, and
$[0.1,0.2]$ for $t_2$.

Our choice of $\ups = 4$ means that
after observing four single failures (and no double failures),
we are prepared to reduce the prior upper fraction of double failures ($0.2$) by half.
Our choice of $\lows = 1$ means that
after observing one double failure (and no single failures),
we are prepared to reduce the prior lower fraction of single failures ($0.8$) by half.

Then, using the 4th order Taylor approximation of $g_2(\vec{\theta})$
as explained in Section~\ref{sec:inference:expectation},
\begin{align}
  \lpr(g_2(\vec{\theta})|\vec{n},\mathcal{H})
  &=
  \min_{(s,\vec{t})\in\mathcal{H}}
  E(g_2(\vec{\theta})|\vec{n},s,\vec{t})
  \\
  &\approx
  \min_{(s,\vec{t})\in\mathcal{H}}
  2\frac{n_2+st_2}{N+s}
  \left(
    1
    -
    \frac{n_2+st_2+1}{N+s+1}
    \left(
      1-\frac{n_2+st_2+2}{N+s+2}
      \left(
        \vphantom{\frac{x}{y}}
      \right.
    \right.
  \right.
  \nonumber
  \\
  & \qquad\qquad\qquad
  \left.
    \left.
      \left.
        1-\frac{n_2+st_2+3}{N+s+3}
        \left(
          1-\frac{n_2+st_2+4}{N+s+4}
        \right)
      \right)
    \right)
  \right)
  \\
  &=0.360
\end{align}
where $N=11$, and $n_2=3$.
A similar expression holds for $\upr(g_2(\vec{\theta})|\vec{n},\mathcal{H})$;
simply replace $\min$ by $\max$:
\begin{equation}
  \upr(g_2(\vec{\theta})|\vec{n},\mathcal{H})=0.410
\end{equation}
Both expressions for lower and upper expecation
are accurate up to the following absolute error:
\begin{equation}
  \max_{(s,\vec{t})\in\mathcal{H}}
  2\prod_{p=0}^5\frac{n_2+st_2+p}{N+s+p}
  =
  0.006
\end{equation}
Concluding,
\begin{multline}
  0.190=(0.360-0.006)\times 0.538
  \\
  \le
  \lpr(q_2|\vec{n},M,T,\mathcal{H},\mathcal{J})
  \le
  \upr(q_2|\vec{n},M,T,\mathcal{H},\mathcal{J})
  \\
  \le
  (0.410+0.006)\times 0.577=0.240
\end{multline}
or in other words, double failures occur at an expected rate that
lies between $0.19$ and $0.24$ per year.

A similar analysis for $q_1$ yields:
\begin{multline}
  0.318=(0.595-0.003)\times 0.538
  \\
  \le
  \lpr(q_1|\vec{n},M,T,\mathcal{H},\mathcal{J})
  \le
  \upr(q_1|\vec{n},M,T,\mathcal{H},\mathcal{J})
  \\
  \le
  (0.643+0.003)\times 0.577=0.373
\end{multline}
or in other words, single failures occur at an expected rate that
lies between $0.318$ and $0.373$ per year.

In this simple example with two redundant components, posterior
imprecision for the single failure rate
is similar to the posterior imprecision for the double failure rate.
This is essentially a special feature of the two component case,
because it must hold that $\theta_1+\theta_2=1$ when $k=2$.
In case of larger $k$, the differences in posterior imprecision between
common-cause failure rates will be considerably larger,
as in the numerical
examples of Section~\ref{sec:idm},
where, for instance, in case of Eq.~\eqref{eq:alphabounds},
$\upr(\theta_j|\vec{n},\mathcal{H}) - \lpr(\theta_j|\vec{n},\mathcal{H})$
ranges from $0.001$ to $0.011$.

\section{Conclusion}
\label{sec:conclusion}

We studied elicitation of hyperparameters for inferences
that arise in the alpha-factor representation of the basic parameter model.
For the hyperparameters of the Dirichlet prior for the alpha-factors,
we argued that bounds, rather than precise values, are desirable, 
due to inferences being strongly sensitive to the choice of prior
distribution, particularly when faced with zero counts.
We concluded that assigning an interval for the learning parameter is especially important.
In doing so, we effectively adapted the
imprecise Dirichlet model \cite{1996:walley::idm} to represent
epistemic uncertainty in the alpha-factors.

For the marginal failure rate, the second part of the model,
we proposed a set of Gamma priors with similar properties as
the set of Dirichlet priors used for the alpha-factors.
As zero counts are generally not an issue for this part of the model,
it may suffice to consider a fixed learning parameter here.

We identified simple ways to elicit information about the hyperparameters,
by reasoning on hypothetical data,
rather than by maximum entropy arguments as done in an earlier study
\cite{1996:atwood,2011:kelly:atwood} on the estimation of alpha-factors.
Essentially, the analyst needs to specify how quickly he is willing to learn
from various sorts of hypothetical data.

Taking everything together,
we arrived at a powerful procedure for analysing
the influence of epistemic uncertainty on
all common-cause failure rates,
the central quantities of interest in the basic parameter model.
As there is no immediate closed-form solution for the expectation
of these failure rates,
we presented an approximation based on Taylor expansion,
and quantified the error of the approximation at any order.

By allowing the analyst to specify bounds for all hyperparameters,
along with clear interpretations of these bounds,
we effectively provided an operational method
for full sensitivity analysis
of common-cause failure rates,
properly reflecting epistemic uncertainty of the analyst
on all levels of the model.
The procedure was illustrated by means of a simple electrical network example,
demonstrating its feasability and usefulness.

In the paper, we chose the sets of hyperparameters to be of
a very specific convex form %.
(Eqs.~\eqref{eq:hyperparams:boxmodel} and~\eqref{eq:gammaprior:box}).
This led to simple calculations (at least for this problem), and
made elicitation fairly straightforward.
Nevertheless, other shapes could still provide a better fit to any given
epistemic information,
and perhaps also have better updating properties.
Such shapes may, however, be more difficult to elicit.
More general shapes for sets of Beta priors
are discussed in \cite{2011:walter:isipta}.
Already for Beta priors, elicitation of these shapes is non-trivial,
and provides an interesting challenge.
We leave a thorough study of such issues, for Dirichlet and Gamma priors, to future work.

Another aspect we neglected in this paper is
the calculation of (imprecise) credible intervals.
We expect that some clever approximation procedure may be needed.

\section*{Acknowledgements}

The first and third author gratefully acknowledge funding from the Durham Energy Institute
(EPSRC Grant EP/J501323/1).
Dana Kelly was originally the lead author for a much earlier incarnation of this paper,
and contributed many of the ideas discussed.
Tragically, he unexpectedly passed away in November 2011.
We express our sincere condolences
to his family, friends, and collaborators.
The authors also thank Simon Blake
for providing fictional yet realistic numbers
for the network reliability example.

\bibliographystyle{amsplainurl}
\bibliography{refs}

\end{document}